\documentclass{aa520}
\usepackage{graphicx,natbib,epsfig}
\bibpunct{(}{)}{,}{a}{}{,}
\begin{document}
\title{A mechanism for parallel electric field generation in the MHD limit:
possible implications for the coronal heating problem in the two stage mechanism}
\author{David Tsiklauri}
\offprints{David Tsiklauri}
\institute{Institute for Materials Research,
University of Salford, Greater Manchester, M5 4WT, United Kingdom.}
\date{Received xxxx / Accepted xxxx}

\abstract{ {\it Context.} Using Particle-In-Cell simulations i.e. in the kinetic plasma description, the 
discovery of a new mechanism of parallel electric field generation was recently 
reported. \\
{\it Aims.} We show that the electric field generation parallel to the uniform unperturbed magnetic field
 can be obtained in a much simpler framework using the ideal magnetohydrodynamics (MHD) description. \\
{\it Methods.} We solve numerically ideal, 2.5D, MHD equations in Cartesian coordinates, 
with a plasma beta of 0.0001 starting from the equilibrium that mimics a 
footpoint of a large curvature radius solar coronal loop or a polar
region plume. On top of such an equilibrium, a purely Alfv\'enic, linearly polarised, plane wave is launched. \\
{\it Results.} In ideal MHD the electric field parallel to the uniform unperturbed magnetic field  
appears due to fast magnetosonic waves which are generated
by the interaction of weakly non-linear Alfv\'en waves with the transverse density inhomogeneity.
In the context of the coronal heating problem  a new  two stage mechanism of plasma heating is
presented by putting emphasis, first, on the generation of parallel electric fields within an  ideal MHD description directly, 
rather than focusing on the enhanced dissipation mechanisms of the Alfv\'en waves and, second, dissipation of these parallel electric 
fields via {\it kinetic} effects. 
It is shown that a single 
 Alfv\'en wave harmonic  with frequency $\nu = 7$ Hz and longitudinal wavelength  $\lambda_A = 0.63$ Mm, for a putative
 Alfv\'en speed of 4328 km s$^{-1}$,
the generated parallel electric field could account for 10\% of the necessary coronal heating requirement. 
It is also shown that the amplitude of the generated parallel electric field 
exceeds the Dreicer electric field by about four orders of magnitude, which implies realisation of the run-away regime
with associated electron acceleration.\\
{\it Conclusions.} We conjecture that  wide spectrum (10$^{-4}-10^3$ Hz) Alfv\'en waves, based on the 
observationally constrained spectrum, could
provide the necessary coronal heating requirement.
The exact amount of energy that could be deposited by such waves through our mechanism of parallel electric field generation can
only be calculated once a more complete parametric study is done. Thus, the "theoretical spectrum" of the energy stored in 
parallel electric fields versus frequency needs to be obtained.

 }
\titlerunning{Parallel electric field generation in the MHD limit...}
\authorrunning{Tsiklauri}
\maketitle

\section{Introduction}

The coronal heating problem, the puzzle of what maintains the solar corona 200 times
hotter than the photosphere, is one of the main outstanding questions in solar physics
(see e.g.  \citet{t05} for a recent brief review on the subject).
A significant amount of work has been done in the context of heating of open
magnetic structures in the solar 
corona \citep{hp83,nph86,p91,nrm97,dmha00,bank00,tan01,hbw02,tn02,tna02,tnr03}.
Historically, all phase mixing studies have been performed in the Magnetohydrodynamic (MHD) approximation;
however, since the transverse scales in the Alfv\'en wave collapse progressively to zero,
the MHD approximation is inevitably violated. 
Thus, \citet{tss05a,tss05b} studied the phase mixing effect in the kinetic regime, using Particle-In-Cell simulations, i.e.
beyond an MHD approximation, where
a new mechanism for the acceleration of electrons
due to the generation of a parallel electric field in the solar coronal context was discovered. 
This mechanism has important implications
for various space and laboratory plasmas, e.g. the 
coronal heating problem and acceleration of the solar wind.
It turns out that in the magnetosphere a similar parallel electric field generation 
mechanism in transversely inhomogeneous plasmas has been reported 
\citep{glm04,glq99}. See also \citet{mgl06} and references therein.

This new mechanism occurs when an Alfv\'en wave moves along the field in
a plasma with a transverse density inhomogeneity. The progressive distortion of the Alfv\'en wave front due to
differences of local Alfv\'en speed then generates the parallel electric field.
In this work we show that the electric field generation parallel to the uniform unperturbed magnetic field
 can be obtained in a much simpler framework
using an ideal MHD description, i.e. without resorting to complicated wave particle interaction effects  such as ion 
polarisation drift and resulting space charge separation, which seems to be the fundamental cause of electron acceleration. 
See also the Discussion section for a more detailed account of what led us to consider the MHD approximation for parallel 
electric field generation.

In this paper, we also explore the implications of this effect for the coronal heating problem.
Preliminary results have been reported elsewhere \citep{t06}.
The present study explores the importance of MHD versus kinetic effects in solar plasmas, which
has recently attracted considerable attention due to apparent difficulties in the coronal heating problem, as well as the natural tendency of
wave heating models to proceed to progressively small spatial scales. 
Also, recent observations have indicated further sub-structuring
of the coronal morphology \citep{ineke2006}, which is relevant 
to our earlier work \citep{t05}.

\section{The model, rationale and main results}

Unlike previous studies \citep{tss05a,tss05b,glm04,glq99}, here we use an ideal MHD description of the problem.
We solve numerically ideal, 2.5D, MHD equations in Cartesian coordinates, 
with a plasma beta of 0.0001 starting from the following equilibrium
configuration: A uniform magnetic field $B_0$ in the $z-$direction penetrates plasma with the density
inhomogeneity across the $x-$direction, which varies according to
\begin{equation}
\rho(x)=\rho_0\left[1+2\left(\tanh(x+10)+\tanh(-x+10)\right)\right]. %%%(1)
\end{equation}
This  means that the plasma density increases from 
some reference background value of $\rho_0$,
which in our case was fixed at 
$\rho_0=2\times10^9 \mu m_p$ g cm$^{-3}$ (with a molecular weight of 
$\mu=1.27$ corresponding to the solar coronal conditions $^1$H:$^4$He=10:1, \cite{a04}
and $m_p$ being the proton mass), to $5 \rho_0$.
Such a density profile across the magnetic field has steep gradients with a half-width of 3 Mm around
$x \simeq \pm 10$ Mm and is essentially flat elsewhere. Such a structure mimics e.g.
the footpoint of a large curvature radius solar coronal loop or a polar
region plume  with the ratio of the density inhomogeneity scale and the loop/plume radius of 0.3, which is
the median value of the observed range 0.15 - 0.5 \citep{g2002}. The above quoted values of coordinate $x$ are in dimensionless
units. We use the following usual normalisation 
$B_{x,y,z}=B_0 \bar{B}_{x,y,z}$,
$(x,y,z)=a_0(\bar{x},\bar{y},\bar{z})$, $t=(a_0/c_A^0)\bar{t}$.
Here we fix $B_0$  to 100 G, and hence the dimensional Alfv\'en speed, 
$c_A^0=B_0/\sqrt{4 \pi \rho_0}$
turns out to be 4328 km s$^{-1}$=0.0144 $c$ ($c$ is the speed of light). 
The reference length $a_0$ was fixed to 
1 Mm, i.e. the dimensionless time of unity corresponds to  0.2311 s.
We usually omit the bar on the top of normalised quantities, hence when
numbers are quoted without units, 
we refer to dimensionless units as defined above.
The dimensionless Alfv\'en speed (normalised to $c_A^0$) is then
$c_A(x)=1 / \sqrt{1+2\left(\tanh(x+10)+\tanh(-x+10)\right)}$.
The simulation domain spans from $-40$Mm to 40 Mm in both $x-$ and $z-$directions
with the density ramp as described above mimicking a footpoint fragment
of a solar coronal loop  or a polar region plume. Our initial equilibrium is depicted in Fig.~1.

\begin{figure}[]
\resizebox{\hsize}{!}{\includegraphics{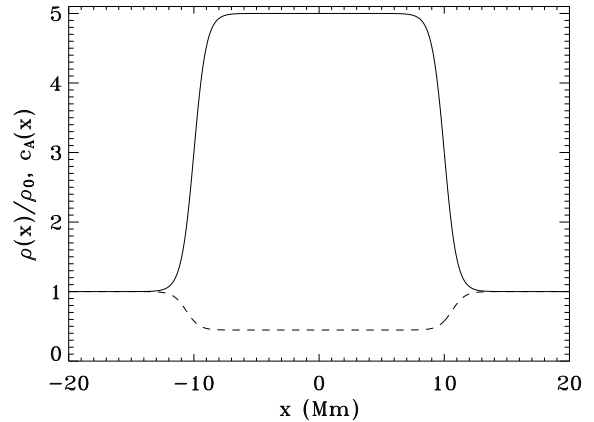}} 
\caption{Dimensionless density (Eq.(1)), solid line, and Alfv\'en speed, dashed line, profiles across the uniform unperturbed magnetic field (i.e. along $x$-coordinate) which is
used as an equilibrium configuration in our model of a footpoint of a solar coronal loop or a polar region plume.}
\end{figure}

The initial conditions for the numerical simulation are
$B_y=A \cos(kz)$ and $V_y=-c_A(x)B_y$ at $t=0$, which means that a purely 
Alfv\'enic, linearly polarised, plane wave is launched travelling in the
direction of positive $z$s. The rest of the physical quantities, 
 $V_x$ and $B_x$ (which 
would be components of fast magnetosonic waves if the medium were totally homogeneous)
and $V_z$ and $B_z$ (the analogs of slow magnetosonic waves) are initially set
to zero. 
The plasma temperature is varied as the inverse of Eq.(1) so that the total pressure always
remains constant.  Boundary conditions used in our simulations are periodic along the $z$- and the zero gradient along the $x$-coordinates.
We fixed the amplitude of the Alfv\'en wave $A$ at 0.05 throughout. This choice makes
the Alfv\'en wave weakly non-linear. Our motivation for this value was two-fold:
First, this was the value used by \citet{tss05b,glm04}, and as one of our
goals here is to show that the parallel electric field generation is also
possible in the MHD limit, we use this value.
 \citet{tss05b} only used one value of $A=0.05$, while
\citet{glm04} used several values of $A$ and showed that weak non-linearity 
(at $A\simeq 0.05$), in addition to the transverse density inhomogeneity, of
course, is a key factor that facilitates parallel electric field generation.
Smaller $A$s reduced the effect of parallel electric field generation.
Second, the observed values of the Alfv\'en waves at heights of
$R=1.04 R_{\sun}\simeq 28$ Mm are about 50 km s$^{-1}$, which for a typical
Alfv\'en speed of 1000 km s$^{-1}$ makes $A$ equal to 0.05.
Such observations (see e.g. \citet{moran01,baner98,doy98}) indicate that 
off limb,  heavy element (e.g. Si VIII) line broadening is
beyond the thermal one, and hence mostly {\it undamped}, outward
propagating Alfv\'en waves travel through the stratified plasma with
increasing amplitude (due to the stratification).

All previous AC-type models of coronal heating focused
on the mechanisms (e.g. phase mixing or resonant absorption) 
that could enhance damping of the Alfv\'en waves.
However, for the coronal value of shear (Braginskii) viscosity,
by which Alfv\'en waves damp of about
$\eta = 1$ m$^2$ s$^{-1}$, typical dissipation lengths ($e$-fold decrease of Alfv\'en wave amplitude
over those lengths) are $\simeq 1000$ Mm. 
Invoking the somewhat ad hoc concept of enhanced (anomalous) resistivity
can reduce the dissipation length to the required value of the 
order of the hydrodynamic pressure scale height $\lambda_T \approx 50$ Mm.
In this light (observation of mostly undamped Alfv\'en waves
and the inability of classical (Braginskii) viscosity to produce a short enough
dissipation length), it seems reasonable to focus rather on the generation
of parallel electric fields which would guarantee plasma heating, should
the energy density of the parallel electric fields be large enough.
 In both cases (AC-type MHD models and our model with parallel electric field generation)
the energy comes from  Alfv\'en waves. Therefore observing undamped  Alfv\'en waves could mean two things: (i) the 
 Alfv\'en waves observed
through Doppler line broadening are low frequency ones, while high frequency  Alfv\'en waves produce heating, hence they do not
 contribute to the observed line broadening. Clearly one cannot observe waves that have already dissipated; (ii) stratification
 (increase of  Alfv\'en wave amplitude with height) could be balanced by the damping.

We emphasise parallel electric fields because in the direction parallel 
to the magnetic field electrons (and protons) are not constrained.
In the direction perpendicular to the magnetic field, 
particles are constrained because of the large classical conductivity 
$\sigma=6 \times 10^{16}$ s$^{-1}$ (for $T=2$ MK corona) and inhibited 
momentum transport across the magnetic field.  The parameter that controls
cross-field transport (flow) is $\sigma$, and not plasma $\beta$ which merely
controls compressibility of the plasma. This is often confused in the literature.

In the ideal, linear MHD limit there are no parallel electric fields associated with the 
Alfv\'en wave. In the single fluid MHD the equation for the electric field
can be obtained by differentiating Ohm's law over time and then using 
Maxwell's equations,  expressing $\eta \vec J$ by $\vec E$ \citep{kt73}:
\begin{equation}
\nabla^2 \vec E - \frac{4 \pi \sigma}{c^2} \frac{\partial \vec E}{\partial t}=
\frac{4 \pi \sigma}{c^2} \frac{\partial }{\partial t} \frac{\vec V \times \vec B}{c}. %%%(2)
\end{equation}
However, in the coronal plasma conductivity is so large that 
a more simple relation (obtained from Eq.(2)) can be used:
\begin{equation}
\vec E = - \frac{\vec V \times \vec B}{c}.%%%(3)
\end{equation}
It is clear from the latter equation that the
parallel electric field is 
\begin{equation}
E_z = - \frac{V_x B_y - V_y B_x}{c}.%%%(4)
\end{equation}
This means that in the considered system 
$E_z$ can only be generated if the initial
Alfv\'en wave ($V_y,B_y$) is able to generate 
a fast magnetosonic wave ($V_x,B_x$),  which, in turn, is generated due to weak non-linearity and
transverse density inhomogeneity.

 \citet{nrm97} have investigated this possibility of growth
of fast magnetosonic waves in a similar physical system, in a context other 
than parallel electric field generation. They
used a mostly analytical 
approach and focused on the early stages of the system's evolution.
Later, the long-term evolution of the fast magnetosonic wave
 generation was studied numerically  in the case
of harmonic \citep{bank00} and Gaussian \citep{tan01} Alfv\'enic
initial perturbations. 
When fast
magnetoacoustic perturbations are initially absent 
and the initial
amplitude of the plane Alfv\'en wave is small, 
the subsequent evolution
of the wave, due to the difference in local Alfv\'en speed
across the $x$-coordinate, leads to the distortion of the wave
front. This leads to the appearance of transverse (with respect to
the applied magnetic field) gradients, which grow linearly
with time. 
\citet{nrm97} have shown that with fairly good accuracy
(which is substantiated by our numerical calculations
presented below) the dynamics of the fast magnetoacoustic
waves can be described by 
$$
\left[\frac{\partial^2 }{\partial t^2} -c_A^2(x)\left(
\frac{\partial^2 }{\partial x^2} + \frac{\partial^2 }{\partial z^2}\right) \right]  V_x=
$$
\begin{equation}
-c_A^2(x)\frac{\partial}{\partial t}\left( B_y \frac{\partial B_y}{\partial x}\right). %%%(5)
\end{equation}
Moreover, for small Alfv\'en wave amplitudes the back-reaction of the generated
fast magnetosonic wave on the Alfv\'en wave can be neglected and
Eq.(5) can be considered as a linear wave-like equation with 
a driver term on the right hand side where $B_y$ is given by the 
travelling wave expression $B_y=A \cos\left(k(z-c_A(x)t)\right)$.
The main negative outcome of the studies with the 
harmonic \citep{bank00} and Gaussian \citep{tan01} Alfv\'enic
initial perturbations
was that despite the power-law growth in time of the driver term,
which implies progressive growth of fast magnetosonic wave
amplitude (at least until $V_x$ and $B_x$ reach the same amplitudes
as $V_y$ and $B_y$ when the the back-reaction can no longer be 
neglected, rendering Eq.(5) invalid), the $V_x$ and $B_x$
amplitudes after rapid initial growth  tend to
saturate due to the destructive wave interference effect \citep{tan01}.
As a self-consistency test, in the next sub-section we present results
of the numerical simulation when the wavenumber of the initial
Alfv\'en wave is $k=1$.  In dimensional units this corresponds to an 
Alfv\'en wave with frequency ($\nu = 0.7$ Hz), i.e. longitudinal wave-numbers $\lambda_A = 6.3$ Mm.

\subsection{The case of an Alfv\'en wave  with $\nu= 0.7$ Hz}

In Fig.~2 we show two snapshots of $V_x$ and $E_z$ (the latter was reconstructed using Eq.(4))
for the case of $k=1$, i.e with $\nu= 0.7$ Hz.
The fast magnetosonic wave ($V_x$) and parallel electric
field ($E_z$) are both generated in the vicinity of the density gradients $x \simeq \pm10$,
eventually filling the entire density ramp. This means that the generated parallel electric
fields are confined by the density gradients, i.e. the solar coronal loop which the considered
system tries to mimic after about 20 Alfv\'en time scales becomes filled with temporally
oscillating parallel electric fields.  Note that this can be a source of polarisation drift of
ions if kinetic effects are considered.

In Fig.~(3) we plot the amplitudes of $V_x \equiv V_x^a$ and $E_z \equiv E_z^a$ which we 
define as the  maxima of absolute values of the wave amplitudes along the $x \simeq \pm10$ line
(which  track the generated wave amplitudes in the strongest density gradient regions) 
at a given time step. Solid lines represent solutions using the linear McCormack code, which
 solves Eq.(5) with the initial conditions as described above, while
dash-dotted lines with open symbols are the solutions using the non-linear 
MHD code Lare2d \citep{tony}. We refer to the McCormack code as linear because
we treat Eq.(5) as a linear wave-like equation with 
a driver term on the right hand side where $B_y$ is given by the 
travelling wave form $B_y=A \cos\left(k(z-c_A(x)t)\right)$.
Here for both runs we use a $2000 \times 2000$ spatial
grid resolution. Note that data with solid lines  is plotted with
 much smaller time steps than those with dash-dotted lines and open symbols,
which are
plotted with the much coarser time step of $\Delta t=1$.
This is due to the fact that Lare2d uses MPI parallelisation and hence 
tracking time evolution of the solution at
a given point of the simulation domain (which
is split into many parts) is a difficult task.
The linear McCormack code is serial, hence we do not encounter such problems.
The two main observations are:

\begin{figure*}
\centering

\epsfig{file=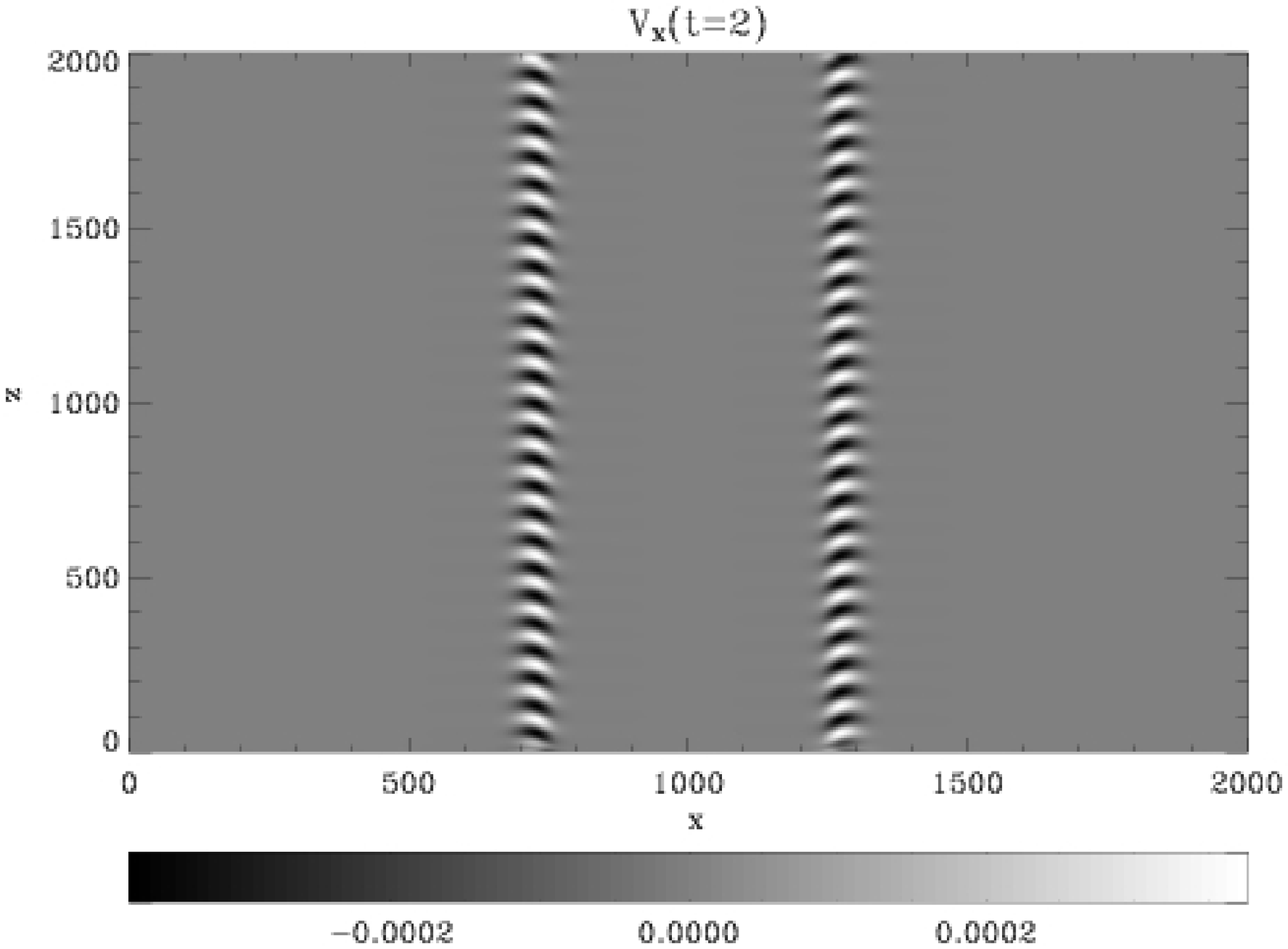,width=6.5cm}
\epsfig{file=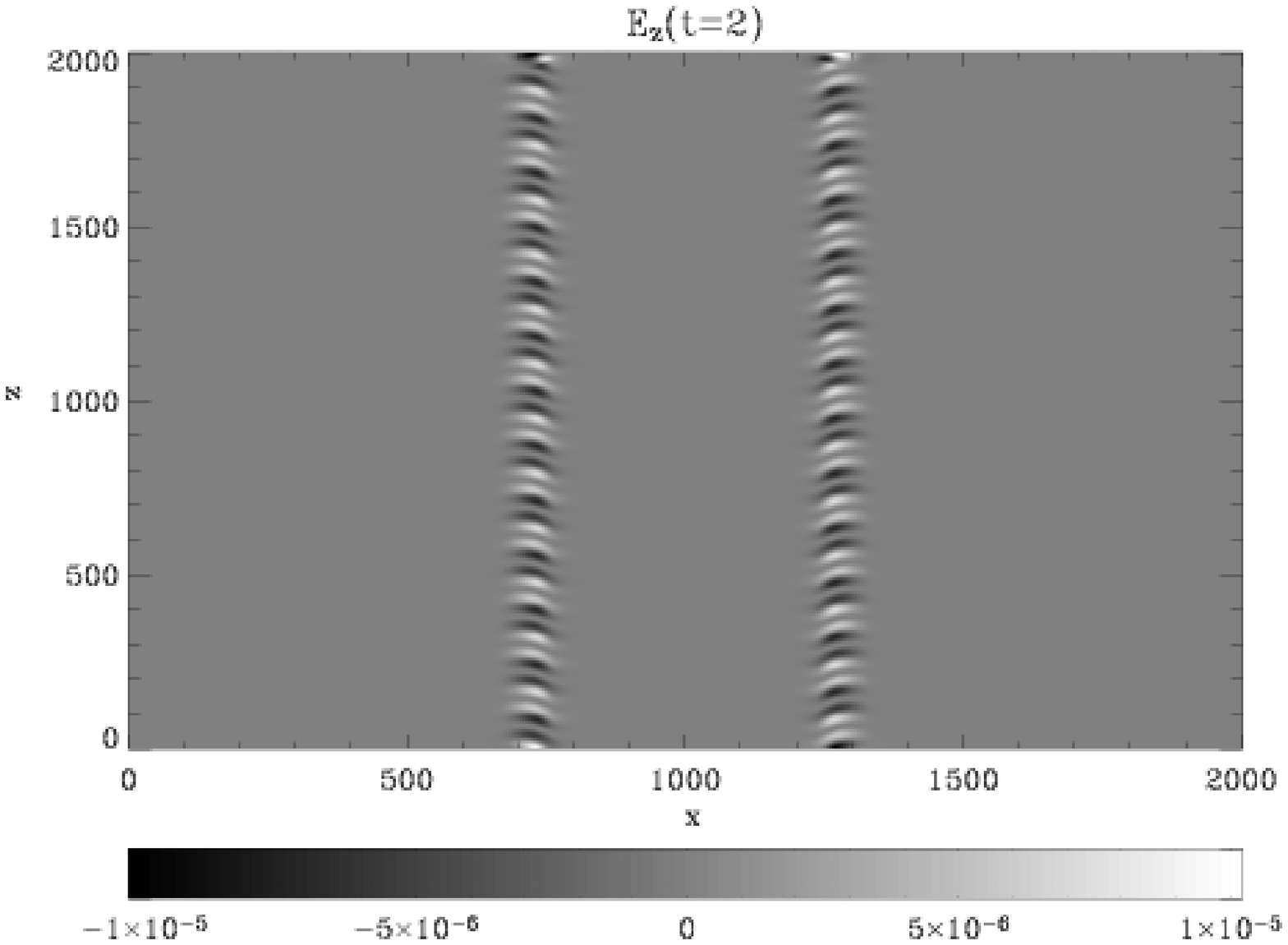,width=6.5cm}
\epsfig{file=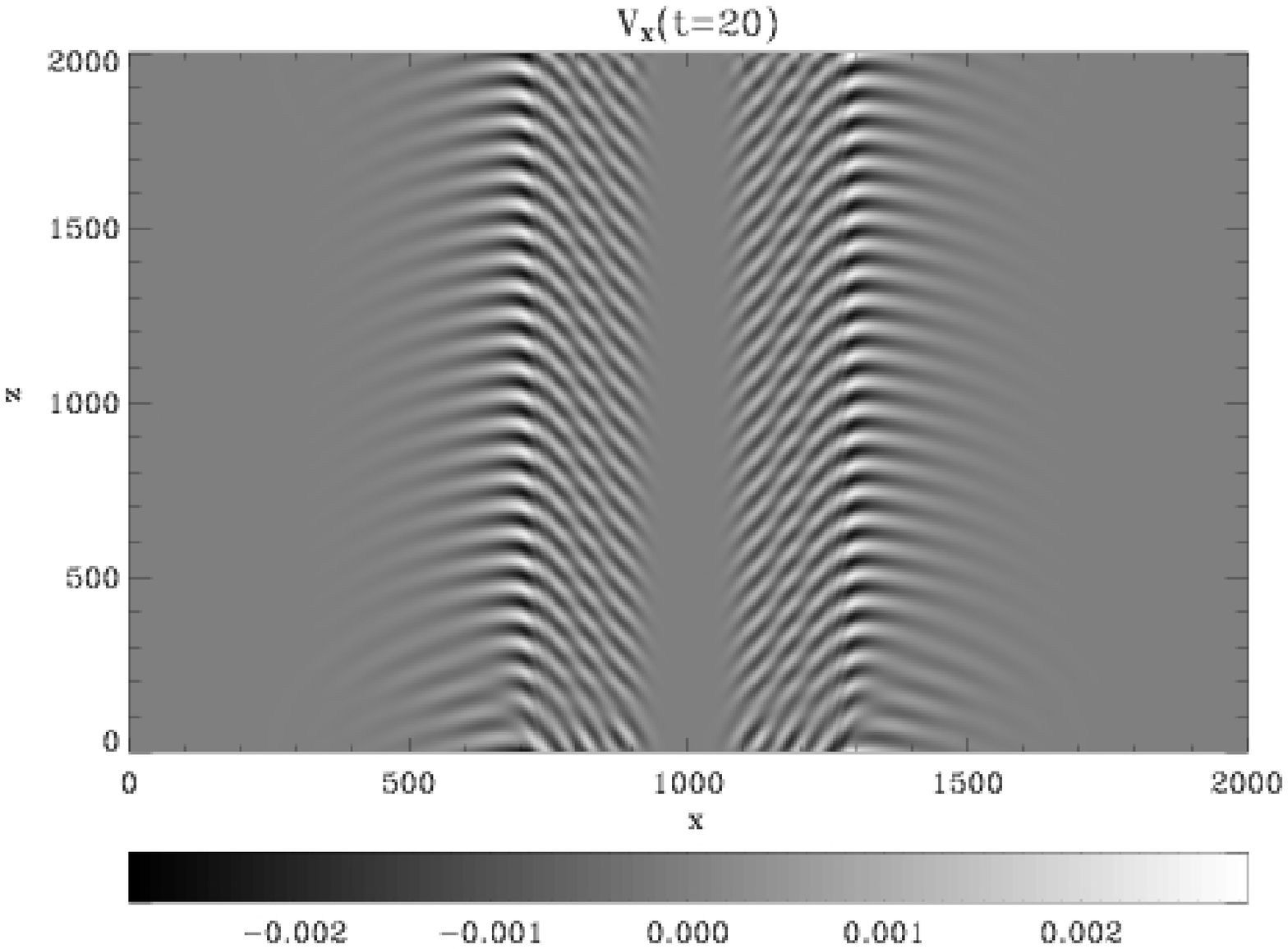,width=6.5cm}
\epsfig{file=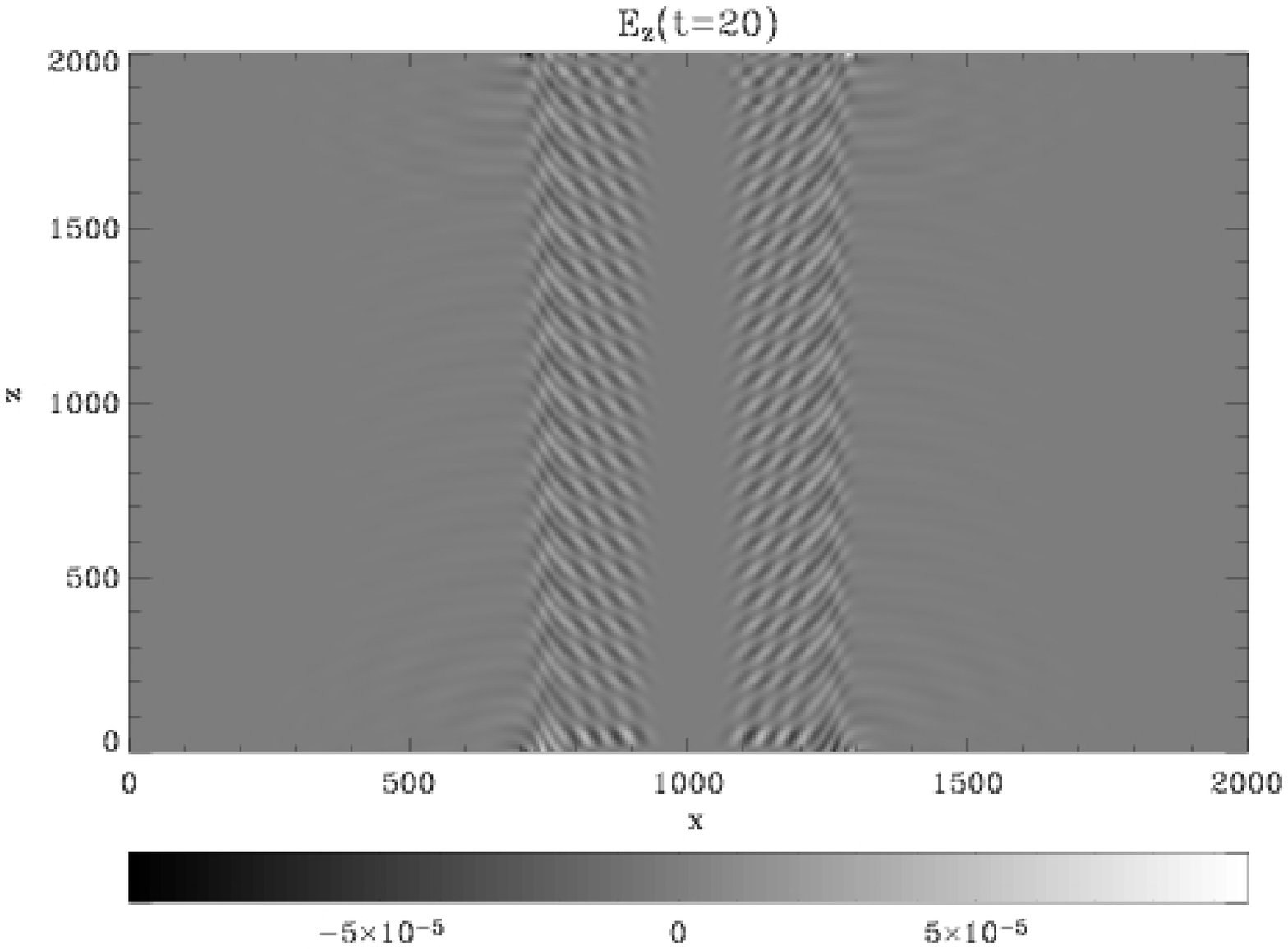,width=6.5cm}

 \caption{Intensity plots of $V_x$ and $E_z$ at $t=2$ and 20 for the case of $k=1$, $\nu = 0.7$ Hz, $\lambda_A = 6.3$ Mm.}
\end{figure*}

\begin{figure}[]
\resizebox{\hsize}{!}{\includegraphics{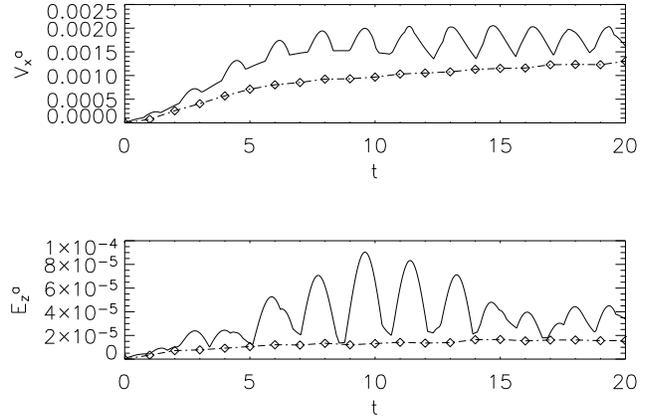}} 
\caption{Time evolution of the amplitudes of $V_x \equiv V_x^a$ and $E_z\equiv E_z^a$.
Solid lines represent solutions using the linear McCormack code, while
dash-dotted lines with open symbols are the solutions using the non-linear 
code Lare2d. Here $A=0.05$, $k=1$, $\nu = 0.7$ Hz, $\lambda_A = 6.3$ Mm, and both runs have $2000 \times 2000$ 
spatial grid resolution.}
\end{figure}

(i) We gather from Fig.~(3) that as in the previous cases
\citep{bank00,tan01} the amplitude of $V_x$ 
initially grows rapidly and then tends to saturate at a level of
$A^2=0.05^2=0.0025$, as one might expect from the weakly
non-linear theory (cf. Eq.(5)).

(ii) The solutions produced by the fully non-linear MHD code and the 
linear McCormack code are similar but not identical. This can be explained
by the fact that the amplitude of $A=0.05$ is now large enough for the
non-linearity effects to be noticeable. Note that when the amplitude
was much smaller, $A=0.001$, the deviations between the two
were much smaller (cf. Fig.(6) from \cite{tan01}).
In the case of $A=0.05$ with Lare2d, weak non-linearity of the phase-mixed
Alfv\'en wave even causes noticeable deviations in the initial 
background density, which is an effect previously discussed  in the
literature.

 Figs.~(2) and (3) corroborate the previous results of 
\citet{bank00} and \citet{tan01}.

\subsection{The case of an Alfv\'en wave  with $\nu= 7$ Hz}

Here we present results for the case of large wave-numbers, $k=10$,   which in dimensional units 
corresponds to an Alfv\'en wave with $\nu = 7$ Hz, and $\lambda_A = 0.63$ Mm.
This is a regime not investigated before.
Fig.~(4) is similar to Fig.(2) but now $k=10$,  with shaded surface plots given instead of intensity plots. This is due to the
fact that spiky data do not appear clearly using intensity plots.
We gather from this graph that similarly to the results of 
\citet{tss05b} and \citet{glm04}, the generated parallel electric field
is quite spiky, but more importantly,  
large wavenumbers i.e. short wavelengths now are able to
significantly increase the amplitudes of 
both the fast magnetosonic waves ($V_x$) and the parallel electric field $E_z$.
This amplitude 
growth is beyond a simple $A^2$ scaling (discussed in the previous sub-section).
 The amplitude growth is presented quantitatively in Fig.~(5).
The amplitude of  $V_x$ now attains values of 0.01, unlike for moderate $k$s.
This boost in amplitude growth can  be explained qualitatively by analysing Eq.(5).
The driver term (right hand side of Eq.(5)) contains spatial derivatives.
Thus, large wavenumbers (i.e. stronger spatial gradients) seem to boost the values of the driver term which in turn
yields larger values for the level of saturation of the $V_x$ amplitude.
In the considered case, $E_z$ now attains values of 0.001.

Since the amplitude of $V_x$ attains a sizable fraction of the Alfv\'en
wave amplitude rendering weakly non-linear theory inapplicable, we do not plot
solutions obtained from the the linear McCormack code.
Instead, to verify the convergence of the solution, we plot the results of the
numerical run with doubled ($4000 \times 4000$) spatial resolution.
The match seems satisfactory, which validates the obtained results.

\begin{figure*}
\centering
\epsfig{file=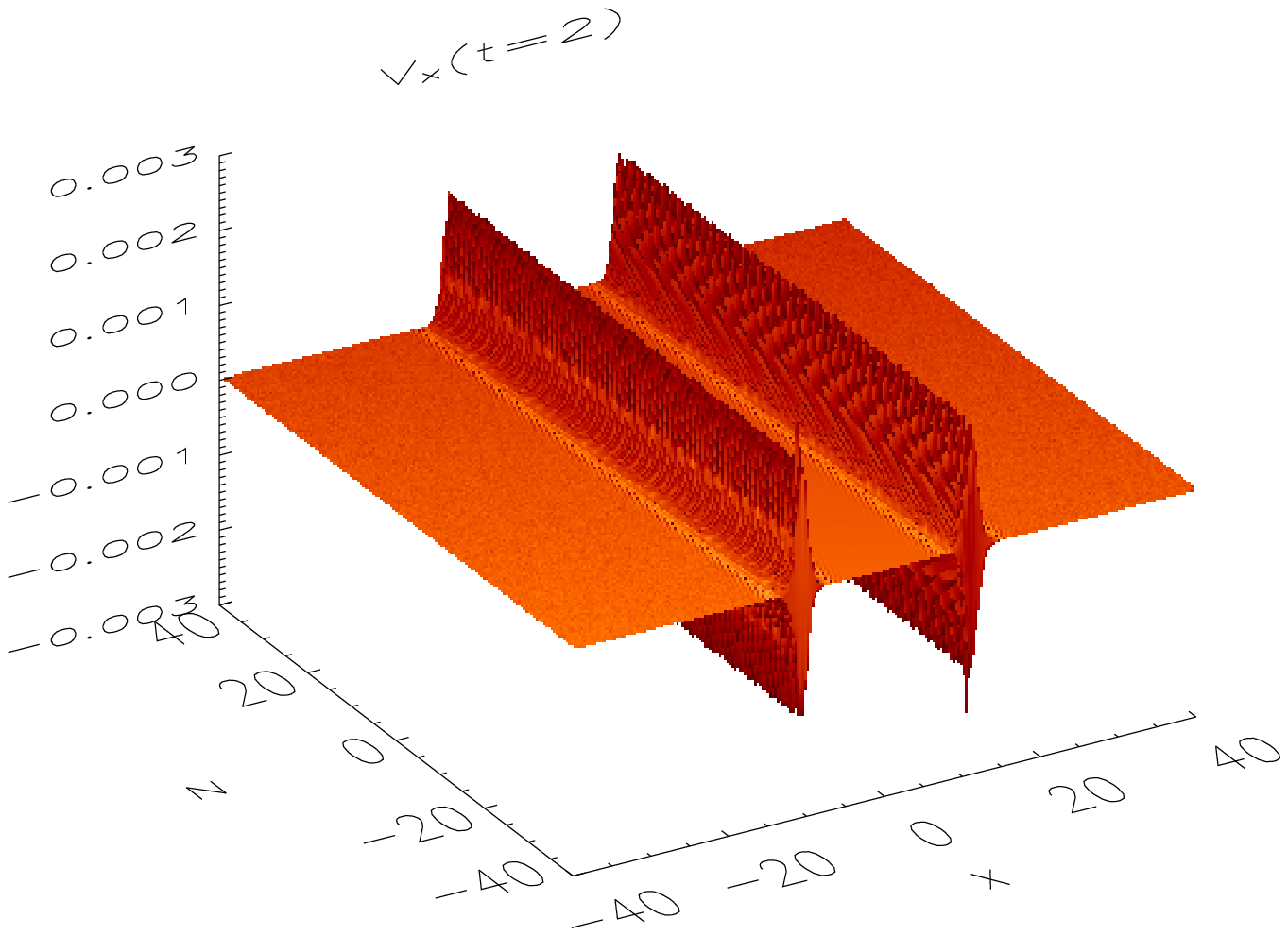,width=6.5cm}
 \epsfig{file=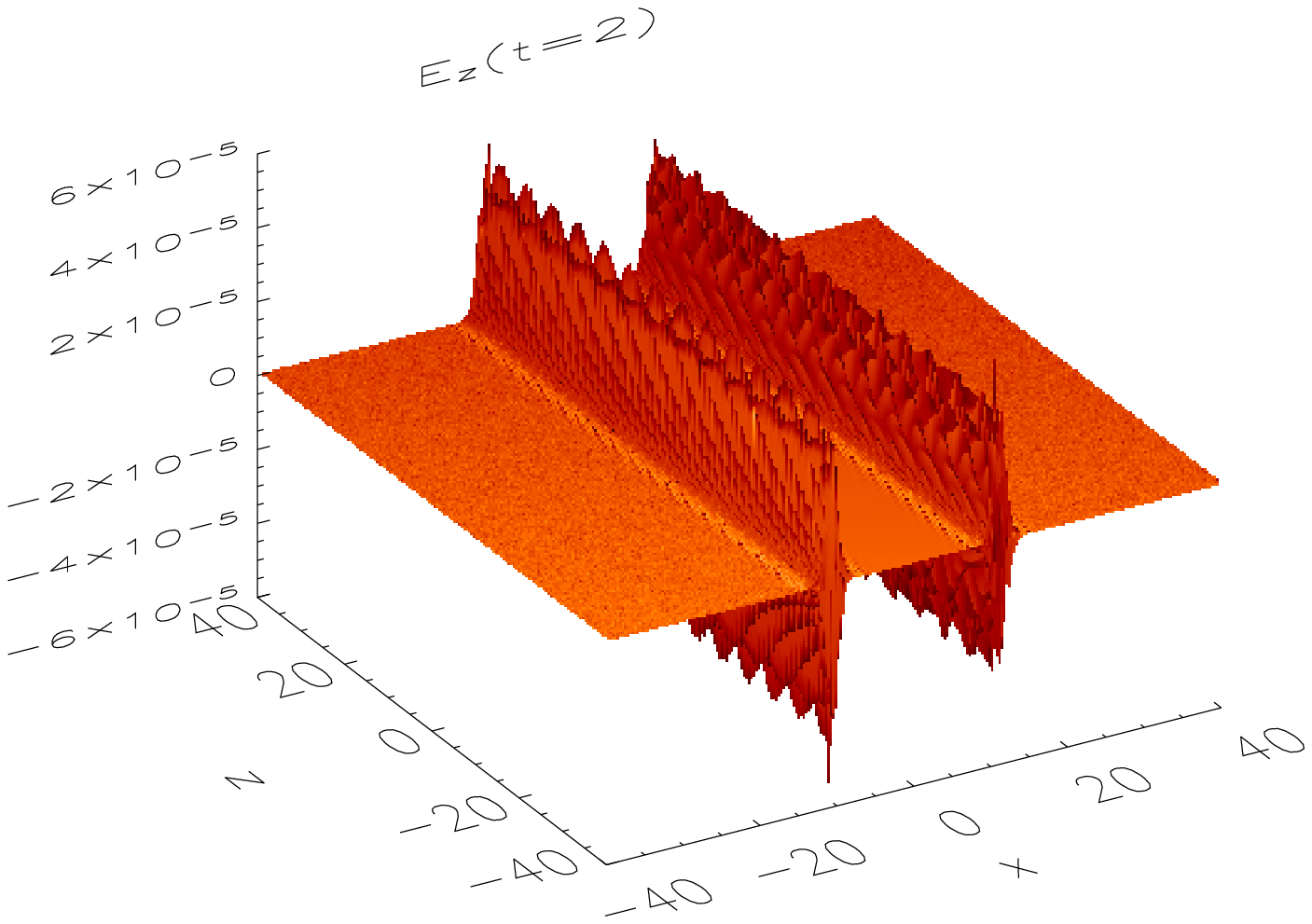,width=6.5cm}
 \epsfig{file=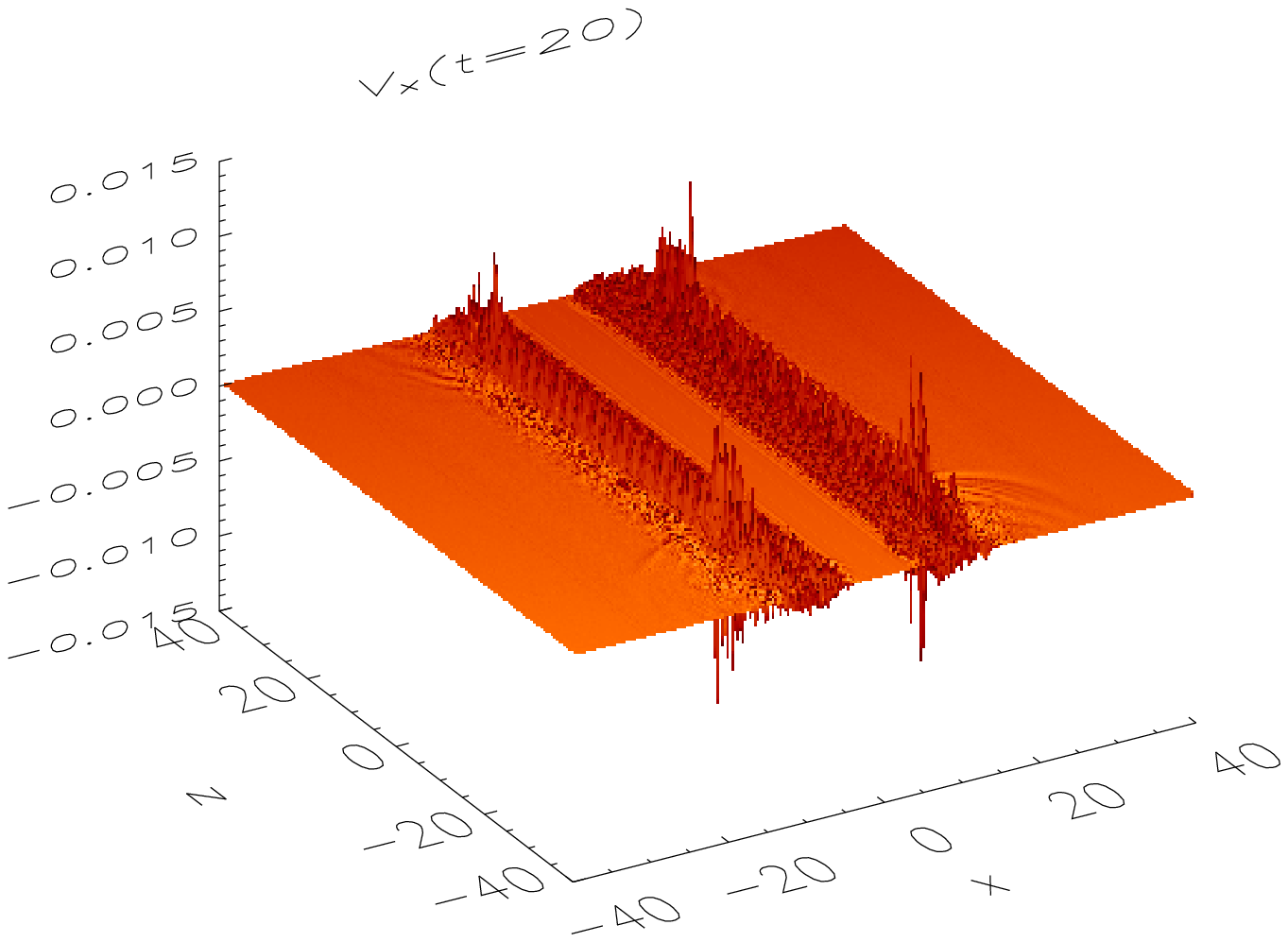,width=6.5cm}
 \epsfig{file=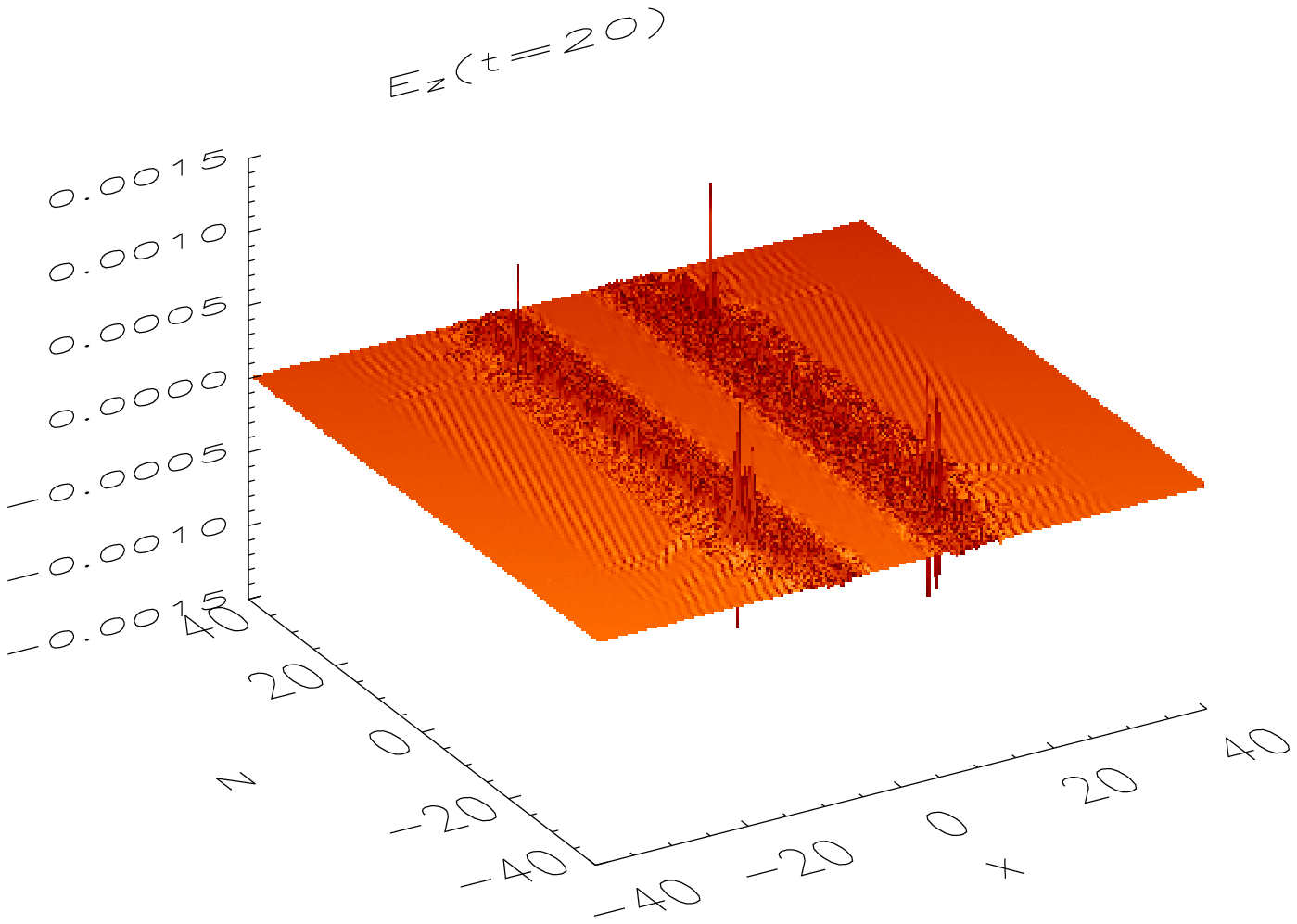,width=6.5cm}
 \caption{Snapshots of $V_x$ and $E_z$ at $t=2$ and 20 for the case of $k=10$, $\nu = 7$ Hz, $\lambda_A = 0.63$ Mm.}
\end{figure*}

\begin{figure}[]
\resizebox{\hsize}{!}{\includegraphics{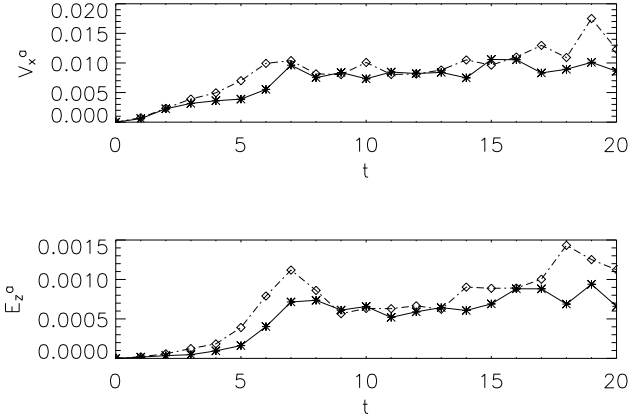}} 
\caption{Time evolution of the amplitudes of $V_x \equiv V_x^a$ and $E_z \equiv E_z^a$.
Solid lines with stars represent solutions using the Lare2d code with $4000 \times 4000$ resolution, while
dash-dotted lines with open symbols are the same but with $2000 \times 2000$ resolution. 
Here $A=0.05$, $k=10$, $\nu = 7$ Hz, $\lambda_A = 0.63$ Mm.}
\end{figure}

\subsection{Application of the results to the coronal heating problem}

It has been known for decades \citep{kis81} that the coronal energy losses that need to be compensated
by some additional energy input, to keep the solar corona at the observed temperatures, are  (in units of erg cm$^{-2}$ s$^{-1}$):
$3\times10^5$ for the quiet Sun, $8\times10^5$ for a coronal hole and $10^7$ for an active region.
\citet{a04} makes similar estimates for the heating flux per unit area (i.e. in erg cm$^{-2}$ s$^{-1}$):
\begin{equation}
F_H=E_H \lambda_T=5 \times 10^3 \left(\frac{n_e}{10^8 {\rm cm}}\right)^2\left(\frac{T}{1 {\rm MK}}\right),
%%%(6)
\end{equation}
where $E_H \approx 10^{-6}$ erg cm$^{-3}$ s$^{-1}$. This yields an estimate of $F_H \approx 2\times10^6$ erg cm$^{-2}$ s$^{-1}$
in an active region with a typical loop base electron number density of $n_e=2\times10^9$ cm$^{-3}$ and $T=1$ MK.

The energy density associated
with the parallel electric field $E_z$ is
\begin{equation}
E_E=\frac{\varepsilon E_z^2} {8 \pi} ,
 \;\;\;
\left[{\rm erg \, cm^{-3} }\right]
%%%(7)
\end{equation}
where $\varepsilon$ is the dielectric permitivity of plasma. The latter can be conveniently deduced
from the expression for the polarisation current (\citet{kt73}, appendix I)
\begin{equation}
\vec J_{\rm p}=n_e q \vec V_{\rm p}=\frac{\rho c^2}{ B^2} \frac{\partial \vec E}{\partial t}
%%%(8)
\end{equation}
and Maxwell's equation written as
\begin{equation}
\nabla \times \vec B=\frac{4 \pi }{c} \left( \vec J + \vec J_{\rm p} \right)
=\frac{4 \pi }{c} \vec J +\frac{1}{c} \frac{4 \pi \rho c^2}{B^2} \frac{\partial \vec E}{\partial t}.
%%%(9)
\end{equation}
The result is 
\begin{equation}
\varepsilon=\frac{4 \pi \rho c^2}{B^2}.
%%%(10)
\end{equation}
The latter formula is different from the usual expression for the 
dielectric permitivity \citep{kt73}:
\begin{equation}
\varepsilon=1+\frac{4 \pi \rho c^2}{B^2},
%%%(11)
\end{equation}
because the displacement current has been neglected in the above treatment, which is a usual
assumption made in the MHD limit. For the coronal conditions ($\rho=2\times10^9 \mu m_p$ g cm$^{-3}$, 
$\mu=1.27$, $B=100$ Gauss) the second term in Eq.(11) is $4.8048 \times 10^3 \approx \varepsilon \gg 1$
which means that for the coronal conditions
in the energy density Eq.(7) the dominant term is 
\begin{equation}
\frac{1}{2}\frac{\rho c^2}{B^2}E_z^2=\frac{1}{2} \rho V_D
%%%(12)
\end{equation}
which is the kinetic energy density associated with the $\vec E \times \vec B$ drift (\citet{sturrok}, page 60).
In Fig.~(5) we saw that electric field amplitude attains a value of $\approx 0.001$. In order to convert
this to dimensional units we use $c_A^0=4328$ km s$^{-1}$ and $B=100$ G and Eqs.(3-4) to obtain
$E_z\approx(c_A^0 B/c)\times 0.001=0.0014$ statvolt cm$^{-1}$ (in Gaussian units).
Therefore the energy density associated
with the parallel electric field $E_z$ (From Eq.(7)) is 
\begin{equation}
E_E=\varepsilon \times 0.0014^2/(8 \pi)=3.7471 \times 10^{-4}
%%%(13)
\end{equation}
$\left[{\rm erg \, cm^{-3} }\right].$

In order to get  the heating flux per unit area for a {\it single harmonic} with frequency
7 Hz, we multiply the latter expression by the 
Alfv\'en speed of 4328 km s$^{-1}$ (because the fast magnetosonic waves ($V_x$ and $B_x$) that propagate across
the magnetic field and associated parallel electric fields ($E_z$) are generated in density gradients by the Alfv\'en waves. 
Hence, the flux is carried with the {\it Alfv\'en} speed)  to obtain
\begin{equation}
F_E=E_E c_A^0=1.62\times10^5\;\;\;
\left[{\rm erg \, cm^{-2} s^{-1}}\right],
\end{equation}
which is $\approx 10$ \% of the
coronal heating requirement estimate for the same parameters made above using Eq.(6).
Note that the latter estimate is for  a {\it single harmonic} with frequency
7 Hz (see the Discussion section for details when a wide spectrum of Alfv\'en waves is considered).

\section{Dissipation of the generated parallel electric fields}
 
We now  discuss how the energy stored in the generated parallel electric field is dissipated. 
We examine the parallel electric field behaviour at a given point in space as a function of
time. In Fig.~(6) we plot the time evolution of $E_z$ at a point $(x=10.68,z=0)$ for the case of $k=1$, $\nu = 0.7$ Hz, and $\lambda_A = 6.3$ Mm.
Choice of this $x$-value is such that it captures parallel electric field dynamics at the strongest density
gradient point (across $x$). 

\begin{figure}[]
\resizebox{\hsize}{!}{\includegraphics{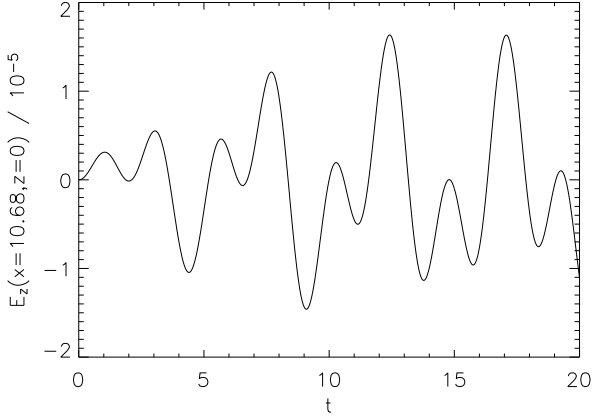}} 
\caption{Time evolution of $E_z$ at a point $(x=10.68,z=0)$. Here $k=1$, $\nu = 0.7$ Hz, and $\lambda_A = 6.3$ Mm.}
\end{figure}

We gather from Fig.~(6) that $E_z$ is a periodic (sign-changing) function that is a mixture
of two harmonics with frequencies $\omega=c_A k$ and $\omega=2 c_A k$. This is due to
 the fact that $E_z$ is calculated using Eq.(4) where $V_y$ and $B_y$ at fixed spatial points
vary in time with frequencies $\omega=c_A k$, while the generated $V_x$ and $B_x$
vary with frequencies $\omega=2 c_A k$ \citep{nrm97}.
Because of the ideal MHD approximation used in this paper, 
the generated electric field cannot accelerate plasma particles or cause Ohmic heating {\it unless kinetic effects are
invoked}. Let us look at Ohm's law for ideal MHD (Eq.(3)) in more detail denoting physical quantities 
unperturbed by the waves
 with subscript 0 and ones associated with the waves by a prime;
initial equilibrium implies $\vec E_0= \vec V_0 =0$ with $\vec B_0 \not = 0$. For the perturbed state 
(with Alfv\'en waves ($V_y$ and $B_y$) launched which generate fast magnetosonic waves ($V_x$ and $B_x$) ) 
we have 
\begin{equation}
\vec E^{\prime}=- {\vec V^{\prime} \times (\vec B_0 +\vec B^{\prime})}/{c}.
\end{equation}      
Note that the projection of $\vec E^\prime$ on the full magnetic field (unperturbed $\vec B_0$ plus the waves $\vec B^\prime$)
is zero by the definition of the cross and scalar product: 
$\vec E^ \prime \cdot (\vec B_0 +\vec B^\prime)/ |(\vec B_0 +\vec B^\prime)| =0$. Physically this means that
in ideal MHD the electric field cannot do any work as it is always perpendicular to the {\it full} (background plus wave) magnetic field.
However the projection of $\vec E^\prime$ on the unperturbed magnetic field $\vec B_0$ is clearly non-zero
\begin{equation}
\vec E^ \prime \cdot \frac{\vec B_0}{ |\vec B_0 |}= E_z=-\frac{\vec V^{\prime} \times \vec B^{\prime} }{c}
\cdot \frac{\vec B_0}{ |\vec B_0 |}
\end{equation}
which exactly coincides with Eq.(4) that was used to calculate $E_z$.
The crucial next step that is needed to understand how the generated electric fields parallel to the uniform unperturbed magnetic field
 dissipate {\it must invoke kinetic effects}. In our two stage model, in the first stage, bulk MHD motion (waves) 
generate parallel electric fields, which as we saw cannot accelerate particles if we describe the plasma in the ideal MHD
limit. \citet{glm04} and \citet{tss05b} showed that when the identical system is
modelled in the kinetic regime particles {\it are} accelerated with such parallel fields.
\citet{glm04} claimed that electron acceleration is due to the polarisation drift.
They showed that once the Alfv\'en wave propagates along the density gradient that is
transverse to the magnetic field, a parallel electric field is generated due to
charge separation caused by the polarisation drift associated with the
time-varying electric field of the Alfv\'en wave. Because the polarisation drift speed is
proportional to the mass of the particle, it is negligible for electrons, hence ions
start to drift. This causes charge separation (the effect that cannot be treated by a MHD
description), which results in generation of parallel electric fields, that in turn accelerate electrons.
As was shown in Fig.(6), our parallel electric field is also time-varying, hence at the second (kinetic) stage the electron
acceleration can proceed in the same manner through ion polarisation drift and charge separation.
The exact picture of the particle dynamics in this case can no longer be treated with MHD and a kinetic
description is more relevant. 
The frequencies considered in the kinetic regime \citep{tss05b}, $\nu=0.3 \omega_{ci}/(2 \pi)=4.6\times10^4$ Hz
and $0.7-7$ Hz (this paper) are different, but we clearly saw that the increase in frequency results in enhancement of the
parallel electric field generation.
Various effects of wave-particle interactions will rapidly dampen the
parallel electric fields on a time scale much shorter than the MHD time scale.
\citet{glm04} clearly demonstrated the role of nonlinearity and kinetic instabilities in the
rapid conversion from the initial low frequency electromagnetic regime to a high frequency
electrostatic one. They identified Buneman and weak beam plasma instabilities in their simulations
(as they studied the time evolution of the system for longer than \citet{tss05b}).
Fig.(11) from \citet{glm04} shows that wave energy is converted into particle energy
on times scales of $10^3 \omega_{pe}^{-1}\approx 4$ Alfv\'en periods.
Perhaps no immediate comparison is possible (because of the different frequencies involved), but still in our case the 
Alfv\'en period when sufficient energy is
stored in the parallel electric field is $1/(7$ Hz)=0.14 s, i.e. wave energy is converted into particle energy in a short time. 

Yet another important observation can be made by estimating the Dreicer electric field \citep{dreicer}.
Dreicer considered dynamics of electrons under the action of two effects: the parallel electric field and friction between electrons and ions.
He noted that the equation describing electron dynamics along the magnetic field can be written as
\begin{equation}
m_e \dot{v_d}=eE-\nu_p^{e/i}m_ev_d
\end{equation}
where $v_d$ is the electron drift velocity and $\nu_p^{e/i}$ is the electron collision frequency and dot denotes time derivative.
When $v_d \ll v_{thermal}$ Eq.(17) in the steady state regime, ($d/dt=0$) allows us to 
derive the expression for Spitzer resistivity. When $v_d > v_{thermal}$, the steady state solution may not apply.
In this case, when the right hand side of Eq.(17) is positive i.e. when $eE>\nu_p^{e/i}m_ev_d$,
we have electron acceleration. This is the so-called {\it run-away regime}.
In simple terms, acceleration due to the parallel electric field leads to an increase in $v_d$;
 in turn, this leads to a decrease in $\nu_p^{e/i}$
because $\nu_p^{e/i}\propto 1/v_d^3$. Thus, when the electric field exceeds the critical value 
$E_d=n_ee^3\ln\Lambda/(8 \pi \varepsilon_0^2k_BT)$ 
(the Dreicer electric field, which is quoted here in its SI form), 
faster drift leads to a decrease in electron-ion friction, which in turn results in even faster drift and
hence the run-away regime is reached. Putting coronal values ($n_e=2\times10^9$ cm$^{-3}$, $T=1$ MK and $\ln\Lambda=17.75$) 
in the expression for the Dreicer electric field we obtain 0.0054 V m$^{-1}$
which in Gaussian units is $1.8\times 10^{-7}$ statvolt cm$^{-1}$.
As can be seen from this estimate, the amplitudes of the parallel electric field obtained in this paper exceeds 
the Dreicer electric field by about four orders of magnitude. This guarantees that the run-away regime would take place
leading to electron acceleration and fast conversion of the generated electric field energy into heat.
In Eq.(17) $E$ is constant, while our $E_z$ at the point of strongest density gradient is
time varying. Hence some modification of the Dreicer analysis is expected.

\section{Discussion}

 After the comment paper by \citet{mgl06} we came to the realisation that the electron acceleration seen in both series of 
works \citep{tss05a,tss05b,glm04,glq99} is a non-resonant wave-particle interaction effect. In works by \citet{tss05a,tss05b}
the electron thermal speed was $v_{th,e}=0.1c$ while the Alfv\'en speed in the strongest density gradient regions
was $v_A=0.16c$; this unfortunate coincidence led us to the conclusion that the electron acceleration by parallel
electric fields was affected by the Landau resonance with the phase-mixed Alfv\'en wave. In works by \citet{glm04,glq99}
the electron thermal speed was $v_{th,e}=0.1c$ while the Alfv\'en speed was $v_A=0.4c$ because they considered a more
strongly magnetised plasma applicable to Earth magnetospheric conditions. 
However, the interaction of the Alfv\'en wave with a transverse density plasma inhomogeneity
when the Landau resonance condition $\omega=k_\parallel c_A(x)$ is met can be quite important for the electron acceleration
\citep{chaston2000,hs76}. \citet{chaston2000} assert that the electron acceleration observed in density cavities in aurorae
can be explained by the Landau resonance of the cold ionospheric electrons with the Alfv\'en wave.
\citet{hs76} also established that at resonance the Alfv\'en wave fully converts into the kinetic  Alfv\'en wave with
a perpendicular wavelength comparable to the ion gyro-radius. We can then conjecture that {\it because of  kinetic Alfv\'en wave front
stretching} (due to phase mixing, i.e. due to the differences in local Alfv\'en speed), 
this perpendicular component {\it gradually realigns} with the ambient magnetic field and hence
creates the time varying parallel electric field component. This points to the importance of the Landau resonance 
for electron acceleration when the resonance condition is met. But as seen in works of  \citet{glm04,glq99},
even when the resonance condition is not met, electron acceleration is still possible.

There were three main stages that lead to the formulation of the present model:

(i) The realisation that the parallel electric field generation (and particle acceleration) is
 a {\it non-resonant} wave-particle interaction effect lead us to the question: 
 could such  parallel electric fields be generated in a MHD approximation? 

(ii) If one considers
{\it non-linear} generation of the fast magnetosonic waves in the transversely inhomogeneous plasma, 
then $\vec E = - (\vec V \times \vec B)/c$ contains a non-zero component
parallel to the ambient magnetic field $E_z = - (V_x B_y - V_y B_x)/c$.  

(iii) From previous studies \citep{bank00,tan01} we knew that the fast magnetosonic waves ($V_x$ and $B_x$) did not grow to
a substantial fraction of the Alfv\'en wave amplitude. However, after reproducing the old parameter regime (k=1, i.e a frequency of 0.7 Hz),
the case of k=10, i.e a frequency of 7 Hz was considered, which showed that  fast magnetosonic waves
and in turn a parallel electric field were more efficiently generated.

 There are two main issues that need to be discussed to address the plausibility of the proposed model:
(i) the plausibility of parameters used in the model and (ii) the relation to the observations.

First, the parameter space of the problem is quite large.
The level at which the fast magnetosonic wave ($V_x$ and $B_x$) and hence the parallel electric field ($E_z$) amplitudes 
saturate depends on several parameters. As indicated previously \citep{bank00,tan01}, this level depends on the strength
of the transverse density gradient and the wavelength (width in the case of a Gaussian Alfv\'enic pulse) of the
Alfv\'en wave. In particular it was found that the stronger density gradients yield lower saturation levels of
the fast magnetosonic wave amplitude (due to the fact that destructive wave interference starts earlier when the
density gradients are strong), while shorter wavelengths (width in the case of the Gaussian Alfv\'enic pulse) of the
Alfv\'en wave generate higher levels (e.g. Figs. 11 and 12 in \citet{tan01}). We have not done a full
parameter space investigation to demonstrate the effect, but instead we fixed the transverse density
gradient guided by observations. In particular, the observed length scale of the density inhomogeneities  in loops vary 
between 0.15 and 0.5 loop radii \citep{g2002}. In our model the length scale of the density inhomogeneity (half-width) is 3 Mm and
the loop radius is 10 Mm (see Fig.~1) which makes the ratio 0.3. This is the median value in the observed range (0.15 -- 0.5).
This eliminates one parameter of variability in the parameter space. 
For the Alfv\'en speed, we used a putative value of  4328 km s$^{-1}$ and performed two numerical runs for two frequencies, 0.7 and 7 Hz.
A full investigation should map a range of frequencies. Such an analysis is pending, but see below for preliminary estimates in the
context of coronal heating.

\begin{figure}[]
\resizebox{\hsize}{!}{\includegraphics{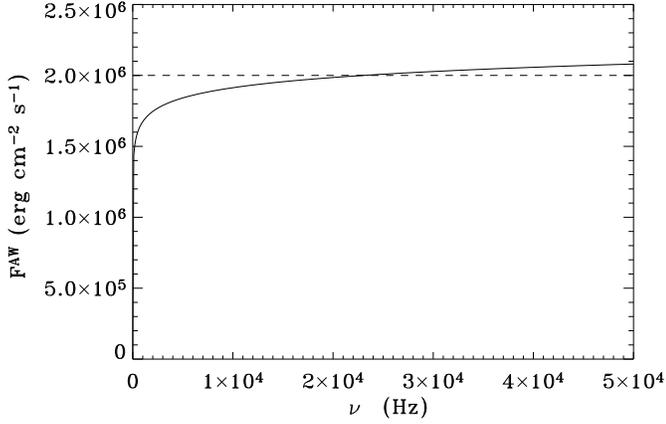}} 
\caption{Plot of the flux carried by  Alfv\'en waves from $10^{-4}$ Hz  up to a frequency $\nu$
versus that frequency as inferred from the
empirically constrained spectrum of \citet{cfk99}. The solid line corresponds to Eq.(18), while the
dashed line shows te coronal heating requirement for a temperature of 1 MK. }
\end{figure}

Second, Alfv\'en waves as observed in situ in the solar wind always appear to be
propagating away from the Sun and it is therefore natural to assume a solar origin for
these fluctuations. However, the precise origin in the solar atmosphere of the hypothetical
source spectrum for Alfv\'en waves (turbulence) is unknown, given the impossibility of remote
magnetic field observations above the chromosphere-corona transition region \citep{vp97}.
Studies of ion cyclotron resonance heating of the solar corona and high speed winds exist which
provide important spectroscopic constraints on the Alfv\'en wave spectrum \citep{cfk99}.
Although the spectrum can and is observed at distances of 0.3 AU, it can be projected back to the base of
corona using empirical constraints, see e.g. the top line in Fig. 5 from \citet{cfk99} (see also the more elaborate model of \citet{cvb05}).
Using the latter figure we can make the following estimates. Let us look at single harmonic, first. At a frequency
of 7 Hz (used in our simulations), the magnetic energy of Alfv\'enic 
fluctuations is $E_\nu^{(\rm 7 \, Hz)} \approx 10^7$ nT$^2$ Hz$^{-1}$. For a single harmonic 
with $\nu=7$ Hz this gives for the energy density 
$E^{(\rm 7 \, Hz)} \equiv \nu E_\nu^{(\rm 7 \, Hz)}/ (8 \pi) \approx 7 \times 10^{-3}$ G$^2/(8 \pi)\approx 2.8 \times 10^{-4}$ 
erg cm$^{-3}$. Surprisingly this {\it semi-observational} value  is quite close to the {\it theoretical} value given by Eq.(13).
As we saw above, such a single harmonic can provide approximately 10 \% of the coronal heating requirement.
Next let us look at how much energy density is stored in the Alfv\'en wave spectrum based on the empirically guided 
top line in Fig. 5 from \citet{cfk99}. 
Their spectral energy density (which they call "power") is approximated by the so-called $1/f$ spectrum, i.e.
$E_\nu=0.6 \times 10^8 / \nu$ nT$^2$ Hz$^{-1}$. In proper energy density units this is $E_\nu=2.4 \times 10 ^{-4} / \nu$
erg cm$^{-3}$ Hz $^{-1}$. Thus, the flux carried by  Alfv\'en waves from say $10^{-4}$ Hz  up to a frequency $\nu$
would be 
\begin{equation}
F^{\rm AW}= \int_{10^{-4} \, {\rm Hz}}^{\nu}E_\nu c_A^0 d \nu= 1.04 \times 10^5 \ln(\nu/10 ^{-4})
\end{equation}
$\left[{\rm erg \, cm^{-2} s^{-1}}\right]$.
It is instructive to look at this equation graphically in Fig.(7): the Alfv\'en wave spectrum from 
$\nu=10^{-4}$ Hz up to about
a few $\times 10^3$ Hz carries a flux that is nearly as much as the coronal heating requirement (Eq.(6)) with $T=1$ MK.
We do not consider higher frequencies because at about $10^4$ Hz, ions become resonant with circularly
polarised  Alfv\'en waves and dissipation proceeds though Landau resonance -- a well-studied mechanism, but quite 
different from our non-resonant mechanism of parallel electric field generation.

There are several possibilities of how this flux carried by Alfv\'en waves (fluctuations) 
is dissipated. If we consider the regime of frequencies up to $10^3$ Hz, the ion cyclotron resonance condition is not
met and hence dissipation is dominated by the mechanism of parallel electric field dissipation formulated in this
paper. However, at this stage it is unclear how much energy could actually be dissipated. This is due to the fact that we
only have two points, 0.7 Hz and 7 Hz, in our "theoretical spectrum". As we saw, a single Alfv\'en wave harmonic with frequency 
7 Hz can dissipate enough heat to account for 10\% of the coronal heating requirement. 
Eq.(18) shows how much flux is carried by the  Alfv\'en waves, while in order to calculate how much of it
is actually dissipated depends on {\it what level $F_E$ from Eq.(14) will attain for each harmonic}. Hence, 
the flux dissipation through our mechanism  would be 
 \begin{equation}
F^{\rm D}= \int_{10^{-4} \, {\rm Hz}}^{\nu}E_\nu c_A^0 D(\nu)d \nu
\end{equation}
where $D(\nu)$ is the "theoretical spectrum" of the energy stored in 
parallel electric fields. For a 7 Hz single harmonic it can be obtained from
\begin{equation}
{\frac{2.4 \times 10 ^{-4}c_A^0 D(\rm 7 \, Hz)}{(\rm 7 \, Hz)}}({\rm 7 \, Hz})=E_E c_A^0
\end{equation} 
with $E_E=3.7471 \times 10^{-4}$ $\left[{\rm erg \, cm^{-3} }\right]$ from Eq.(13) rendering
$D(\rm 7 \, Hz)=1.56$. For a 0.7 Hz single harmonic $E_E$ would be different because as can be seen from Fig.(3) $E_z\approx 3 \times 10^{-5}$ (as opposed to 
$E_z=0.001$ for $\nu=7$ Hz). Since $E_E$ scales as $E_z^2$, then $D(\rm 0.7 \, Hz)=1.56 \times (3 \times 10^{-5}/0.001)^2 \approx 1.4 \times 10^{-3}$.
More numerical runs are needed to fully map $D(\nu)$. This is deferred until further work is done.

A clear distinction should be made between our model and the ones that use ion cyclotron resonance damping of Alfv\'en waves.
Low frequency (0.001-0.1 Hz) Alfv\'en waves dominate the power spectrum of fluctuations in the solar wind at and beyond 0.3 AU.
These waves are able to transport a significant amount of energy \citep{cfk99}. On the contrary, high frequency (10-10000 Hz) Alfv\'en waves 
are known to damp more easily than their low frequency counterparts, but they are not expected to contain much power.
This is problematic for ion cyclotron resonance damping models because Landau resonance of ions occurs at high frequencies (few $10^4$ Hz).
Our model on the contrary is of a non-resonant nature. Thus, it is hoped that it can provide enough heating once the frequency range
10$^{-4}-10^3$ Hz at the base of corona is mapped numerically (i.e. once the level attained by the parallel electric field amplitudes for each frequency is determined).
The proposed wide spectrum idea for Alfv\'en waves is not as "theoretical-rather-than-demonstrable"
as the one proposed by \citet{tn01} for the slow magnetosonic waves. One could question the validity of the wide spectrum idea
for the slow magnetosonic waves in coronal loops as predominantly single
harmonics (with periods of 3 and 5 minutes, etc.) are observed. At the same time it is not possible to observe 
the high frequency waves that  are dissipated already.
However, in the case of Alfv\'en waves, the  presence of a wide spectrum with a frequency range of
10$^{-4}-10^3$ Hz at the base of corona, which is 
actually observed at 0.3 AU,  seems to be more likely. The possibility of meeting the full coronal heating requirement with the 
wide spectrum Alfv\'en waves 
via the proposed two stage mechanism of parallel electric field generation needs further investigation.

\begin{acknowledgements}
The author acknowledges support from the Nuffield Foundation (UK) through an award to newly 
appointed lecturers in Science, Engineering and Mathematics (NUF-NAL 04), from the 
University of Salford Research Investment Fund 2005 grant, PPARC (UK) standard grant and 
use of the E. Copson Math cluster funded by PPARC 
and the University of St. Andrews.  The author would like to thank E.R. Priest and K.G. McClements for
useful discussions and criticism and the anonymous referee for useful comments and suggestions.
\end{acknowledgements}

\bibliography{ms4816}% Produces the bibliography via BibTeX.
\end{document}